\documentclass[conference, 10pt]{IEEEtran}

\usepackage{listings}
\usepackage{graphicx}
\usepackage{xcolor}
\usepackage{tabu}
\usepackage{tabularx}
\usepackage{svg}
\usepackage{subcaption}
\usepackage{url}
\usepackage{float}
\usepackage{hyperref}
\newcolumntype{Y}{>{\centering\arraybackslash}X}

\lstdefinestyle{basic}{
  extendedchars     = {true},
  inputencoding     = {utf8},
  basicstyle        = {\ttfamily \footnotesize},
  keywordstyle      = {\rmfamily \bfseries},
  commentstyle      = {\rmfamily \itshape},
  tabsize           = {2},
  flexiblecolumns   = {false},
  frame             = {single},
  showstringspaces  = {false},
  breaklines        = {true},
  breakatwhitespace = {true}
}

\lstdefinelanguage{PS}
{
  morekeywords = {
    true, false, <nullptr>
  },
  sensitive    = true,
  morecomment  = [l]{---}
}

\lstdefinestyle{PS}{
  style    = {basic},
  frame    = {none},
  language = {PS}
}

\lstdefinelanguage{Kotlin}{
  keywords = {
    package, as?, typealias,
    this, super, val, var,
    fun, for, null, true,
    false, throw,
    return, break, continue, object,
    if, try, else, while,
    do, when, class, interface,
    enum, companion, override, public,
    private, get, set, import,
    abstract, vararg, expect, actual,
    where, suspend, data, internal,
    dynamic, final, by
  },
  keywordstyle = {\bfseries},
  ndkeywords = {
    @Deprecated, @JvmName, @JvmStatic, @JvmOverloads,
    @JvmField, @JvmSynthetic, Iterable, Int,
    Long, Integer, Short, Byte,
    Float, Double, String, Runnable,
    Array
  },
  ndkeywordstyle = {\bfseries},
  emph = {
    println, return@, forEach, map,
    mapNotNull, first, filter, firstOrNull,
    lazy, delegate
  },
  emphstyle       = {},
  identifierstyle = \color{black},
  sensitive       = true,
  commentstyle    = {\color{gray}\ttfamily},
  comment         = [l]{//},
  morecomment     = [s]{/*}{*/},
  stringstyle     = {\ttfamily},
  morestring      = [b]",
  morestring      = [s]{"""*}{*"""},
}

\begin{document}

\newcommand{\gpt}{ChatGPT-4}
\newcommand{\gpto}{ChatGPT-4o}
\newcommand{\codellama}{Code Llama 70b}
\newcommand{\llama}{Llama Medium}

\newcommand{\gitbug}{GitBug Java}
\newcommand{\testspark}{\texttt{TestSpark}}
\newcommand{\evosuite}{\texttt{EvoSuite}}
\newcommand{\kex}{\texttt{Kex}}
\newcommand{\testsparkgpto}{\testspark{}\texttt{-\gpto{}}}

\newcommand{\intellij}{IntelliJ IDEA}
\renewcommand{\rq}[1]{RQ\textsubscript{#1}}
\newcommand{\sbst}{SBST}
\newcommand{\llm}{LLM}
\newcommand{\llms}{\llm{}s}
\newcommand{\cut}{CUT}
\newcommand{\cuts}{\cut{}s}
\newcommand{\projectUnderTest}{PUT}
\newcommand{\projectsUnderTest}{\projectUnderTest{}s}
\newcommand{\llmBased}{\llm{}-based}
\newcommand{\se}{symbolic execution}
\newcommand{\seBased}{\se{}-based}
\newcommand{\sloc}{SLOC}

\title{Test Wars: A Comparative Study of SBST, Symbolic Execution, and LLM-Based Approaches to Unit Test Generation}

\author{\IEEEauthorblockN{1\textsuperscript{st} Azat Abdullin}
\IEEEauthorblockA{
\textit{JetBrains Research, TU Delft}\\
Amsterdam, The Netherlands\\
azat.abdullin@jetbrains.com
}
\and
\IEEEauthorblockN{2\textsuperscript{nd} Pouria Derakhshanfar}
\IEEEauthorblockA{
\textit{JetBrains Research}\\
Amsterdam, The Netherlands \\
pouria.derakhshanfar@jetbrains.com}
\and
\IEEEauthorblockN{3\textsuperscript{rd} Annibale Panichella}
\IEEEauthorblockA{
\textit{TU Delft}\\
Delft, The Netherlands \\
a.panichella@tudelft.nl}
}

\maketitle

\begin{abstract}
Generating tests automatically is a key and ongoing area of focus in software engineering research. The emergence of Large Language Models (\llms{}) has opened up new opportunities, given their ability to perform a wide spectrum of tasks. However, the effectiveness of \llmBased{} approaches compared to traditional techniques such as search-based software testing~(\sbst{}) and symbolic execution remains uncertain.
In this paper, we perform an extensive study of automatic test generation approaches based on three tools: \evosuite{} for \sbst{}, \kex{} for \se{}, and \testspark{} for \llmBased{} test generation. We evaluate tools' performance on the \gitbug{} dataset and compare them using various execution-based and feature-based metrics.
Our results show that while \llmBased{} test generation is promising, it falls behind traditional methods in terms of coverage. However, it significantly outperforms them in mutation scores, suggesting that \llms{} provide a deeper semantic understanding of code. \llmBased{} approach also performed worse than \sbst{} and \seBased{} approaches w.r.t. fault detection capabilities. Additionally, our feature-based analysis shows that all tools are primarily affected by the complexity and internal dependencies of the class under test~(\cut{}), with \llmBased{} approaches being especially sensitive to the \cut{} size.
\end{abstract}

\begin{IEEEkeywords}
automatic test generation, symbolic execution, concolic testing, large language models, search-based software testing
\end{IEEEkeywords}

\section{Introduction}
\label{section:intro}

Software quality assurance is a critical aspect of the software development process, as errors in the software may lead to fatal consequences. Software testing remains one of the most widespread software quality assurance methods~\cite{ammann2017introduction}, manual testing being the most popular type of software testing~\cite{bellerfse2015}. Yet, automated solutions are beneficial since manual testing is a complex and time-consuming task~\cite{manualTesting}. As a result, researchers have proposed various methods for generating test cases automatically: search-based software testing~(\sbst{})~\cite{evosuite, pynguin}, \se{}~\cite{klee}, concolic testing~\cite{kex}, etc. These techniques have advanced significantly, achieving high code coverage~\cite{jahangirova2023sbft}, leading to fewer smells than manually written test cases~\cite{panichella2022test}, and detecting unknown bugs~\cite{fraser20151600}. Adoption of automated test generation tools in industry is still limited, despite these advancements~\cite{TestGenerationInIndustry}.

Large Language Models (LLMs) offer new possibilities in software engineering, showing effectiveness in tasks like code completion~\cite{LLM4CodeCompletion}, code understanding~\cite{LLM4CodeUnderstanding}, and code generation~\cite{LLM4CodeGeneration}. More recently, \llms{} have been applied to automatic test case generation~\cite{llm4testGen}, leveraging their ability to understand natural language and code context. However, \llms{} face challenges such as limited context windows and hallucinations~\cite{LlmHallucination}. While some empirical studies suggest that \llms{} are useful for test generation, they often fall short compared to traditional approaches like \sbst{}, particularly when handling large classes~\cite{LLMvsSBST}.

We identify three critical limitations in existing comparisons between \llmBased{} and traditional approaches. \textit{Limitation 1: Benchmark and data contamination}.
Existing \llmBased{} approaches~\cite{bhatia2024unit, chen2024chatunitest, gu2024testart, tufano2020unit,LLMvsSBST, yuan2024evaluating} have been evaluated on datasets like Defects4J~\cite{gu2024testart, Defects4j} or popular GitHub projects~\cite{bhatia2024unit, chen2024chatunitest}, which are part of the \llms{} (pre)training data, introducing risks of data leakage~\cite{lee2024github, sallou2024breaking}. Therefore, the evaluation should be conducted using projects from different sources~\cite{sallou2024breaking} or commits and defects discovered and fixed after the release date of the \llms~\cite{lee2024github}. Recent works~\cite{siddiq2024using, LLMvsSBST} partially tackle this issue using the  SF110 dataset or its extensions. However, this dataset has been extensively used to compare and improve \sbst{} tools such as \evosuite{}, thus introducing a potential positive bias towards them. 



\textit{Limitation 2: No comparison with symbolic execution.} \sbst{} approaches are efficient and effective \cite{SBSTSurvey, sbst} but they are not the only stat-of-the-art technique for generating unit tests. 
Symbolic execution is the first technique ever used in the literature to generate test inputs~\cite{ramamoorthy1976automated}. These techniques use constraint solvers to generate inputs that satisfy the conditions in the code. Still, they might struggle to satisfy conditions and large classes due to the path explosion problem~\cite{SymbolicExecutionChallenges}. Existing studies with \llms{} did not consider these techniques.

\textit{Limitation 3: Lack of statistical analysis and repetitions.}
Existing studies do not fully account for the non-determinism of \llms{}, which may produce different outputs when queried with the same prompt. This issue is seldom acknowledged~\cite{LLMvsSBST} but is yet to be addressed. In fact, \llms{} have been executed only once (one seed/session), not following existing guidelines on how to assess randomized tools~\cite{GuideToStatisticalTests, sallou2024breaking} statistically.


In this paper, we conduct an extensive comparative study that addresses the three limitations discussed above. We compare three cutting-edge tools for unit test generation, namely \evosuite{} (\sbst{} tool), \kex{} (\se{} tool), and \testspark{}. The latter tool has been configured with different \llms{}: \gpt{}, \gpto{}, \llama{}, \codellama{} as we investigate the difference in performance between alternative \llms{} when given with the same prompt and source code information. 
We use \gitbug{} as the benchmark, a dataset of recent Java commits and bugs published in 2024. This dataset is designed to address data leakage issues and 
has not been used in prior \sbst{} studies.

We assess each tool's performance based on execution metrics (e.g., code coverage) and feature-based metrics (e.g., complexity of the class under test). Finally, we run each tool ten times with different seeds (and independent sessions for \testspark{}) and base our analysis on sound statistical analysis as suggested in the literature~\cite{GuideToStatisticalTests, sallou2024breaking}. Our study has two primary goals: (1) to compare the performance of \llms{}, \sbst{}, and \se{} in generating unit tests, and (2) to analyze the strengths and weaknesses of each approach, which could inform the development of hybrid techniques.

Overall, our results suggest that \gpto{} has the best potential in automatic test generation among all \llms{}. However, according to statistical tests, \testsparkgpto{} still performed worse than traditional methods. We further provide insights into the characteristics of the program under tests that seem to impact the performance of the different tools, shedding light on their advantages and disadvantages.

Our main contributions can be summarized as follows:
\begin{itemize}
    \item Open source extendable automatic test generation assessment pipeline\footnote{\url{https://github.com/plan-research/tga-pipeline}}.
    \item An extensive analysis of three automatic test generation techniques --- \sbst{}, \seBased{}, and \llmBased{} --- in terms of achieved compilation rate, coverage, and mutation score of generated tests.
    \item A co-factor analysis between the automatic test generation techniques performance and the features of the code under test: complexity, size, language features, etc.
\end{itemize}

This paper is organized as follows. Section~\ref{section:related} gives an overview of the prior work in our area. The design and implementations of the presented test generation pipeline are described in Section~\ref{section:pipeline}. Section~\ref{section:experimentalsetup} describes all the details of the experimental setup, while~\ref{section:evaluation} presents the evaluation results. Finally, Section~\ref{section:discussion} discusses the most exciting results found during the evaluation, and Section~\ref{section:threats} discusses potential threats to validity, while Section~\ref{section:conclusion} concludes the paper.

\section{Related work}
\label{section:related}

\subsection{Traditional test generation approaches}
Search-based software testing~(\sbst{})~\cite{sbst} is one of the most popular and effective automatic test generation methods. The main idea of \sbst{} is to use meta-heuristic optimization methods like genetic algorithms to generate test cases/suites. \evosuite{}~\cite{evosuite} and Pynguin~\cite{pynguin} are two of the most popular \sbst{} tools developed for Java and Python, respectively. They have won the SBFT/SBST unit test competitions for several years~\cite{erni2024sbft, SBST22Competition, SBFT23Competition, SBST2021}.

Symbolic execution~\cite{SymbolicExecution} is a software analysis technique that abstractly executes the target program while substituting the input values with symbolic variables. Automatic test generation is one of the main applications of symbolic execution. Concolic testing~\cite{ConcolicTesting} is a technique that tries to solve some of the problems of symbolic execution by combining it with concrete execution. Concolic testing has been proven to be very effective in automatic test generation by tools like Klee~\cite{klee}, \kex{}~\cite{kex}, UtBot~\cite{UtBot}, etc.

Both \sbst{} and symbolic execution have proven to be effective and reliable test generation approaches throughout the years~\cite{SEandSBST}. This work focuses on Java test generation and includes evaluation of both \evosuite{} and \kex{}.

\subsection{\llmBased{} test generation approaches}
\testspark{}~\cite{testSpark}, ChatUniTest~\cite{chatUniTest} and TestPilot~\cite{testPilot} are examples of \llmBased{} test generation tools. TestPilot is a standalone tool for JavaScript programs, while TestSpark and ChatUniTest work with Java and provide \intellij{} plugins for better user experience. These tools are similar in their approach to test generation, which mainly consists of three steps: (1) prompt collection, (2) \llm{} request, and (3) a feedback loop that tries to improve the initial \llm{} response. 

SymPrompt~\cite{symPrompt} introduces an \llmBased{} test generation that focuses on generating tests at the method level rather than for classes. It uses code-aware prompts to target specific paths within the method under test. While effective, this approach requires more time and incurs additional \llm{} requests, making it slower than other \llmBased{} tools. Similarly, AthenaTest~\cite{tufano2020unit}  generates tests for individual (focal) methods. It does not include a feedback loop for the LLM, and it has been evaluated using Java static methods. We note that method-level tools have limited applicability for more complex, object-oriented scenarios involving inheritance and stateful classes. 

AID~\cite{aid} is a \llmBased{} approach that targets bug detection. This approach focuses on test generation for coding problems and stands out because it uses program specification as a part of \llm{} prompt and requests \llm{} to provide a test input generator script instead of the actual tests. This approach has shown promising results; however, it focuses on a specific use case, not general-purpose unit test generation.

Our study focuses on unit test generation in the "traditional" setting ~\cite{SBFT23Competition}, where all the competitors have the same limited time budget and generate class-level tests. In that setting, \llmBased{} tools like \testspark{} and ChatUniTest are the most suitable. We consider \testspark{} for our evaluation since it can support various LLMs, such as ChatGPT, which is very common for many \llmBased{} test generation approaches~\cite{bhatia2024unit, chen2024chatunitest, siddiq2024using}. Additionally, the tool can set a strict time window, automatically compile the generated tests, and integrate a feedback loop that can refine the responses of LLM. Furthermore, since \testspark{} utilizes \intellij{} in headless mode, all the context collection and code inspection features available in this IDE can be employed for prompt generation using the command line interface. As a result, we can enhance the context provided for prompt generation in our evaluation process.

\subsection{Hybrid test generation approaches}
A group of approaches also combines \llmBased{} test generation with traditional automatic test generation methods. CodaMosa~\cite{CodaMosa} is one of the first successful attempts that combined EvoSuite with LLMs. The core idea of CodaMosa is to fall back to \llmBased{} test generation when the search-based approach reaches its coverage plateau. CoverUp~\cite{CoverUp} is an extension of CodaMosa that improves the \llm{} feedback cycle by enhancing it with coverage information. 
However, our paper mainly focuses on studying the strengths and weaknesses of individual approaches, which can serve as the foundation for developing future hybrid strategies. Therefore, we exclude hybrid approaches from our work. Based on our findings, as future work, we plan to design new strategies to combine different tribes of AI for unit test generation and then use the existing hybrid test generation approaches as baselines.

\subsection{Existing comparative studies}
Recent attempts to perform a comparative study of \llmBased{} test generation with traditional approaches have been presented in the literature~\cite{LLMvsSBST,siddiq2024using, gu2024testart,yuan2024evaluating}. These studies focus on evaluating the test-generating capabilities of ChatGPT 3.5 in comparison with \evosuite{}. Tang et al.~\cite{LLMvsSBST} also highlight the strengths and weaknesses of these two approaches. However, the studies above have clear limitations, which are also discussed in the introduction:
\begin{itemize}
    \item All studies related to ChatGPT focus on version 3.5, which more recent versions have replaced. Besides, they do not compare different potential \llms{} to use.
    \item These studies evaluate \llms{} using either the SF10~\cite{SF100}, Defects-4j~\cite{Defects4j} datasets, or popular projects from GitHub. Recent studies highlight that Defects-4j and many existing GitHub projects are an integral part of the model (pre)training dataset, leading to data leakage~\cite{lee2024github, sallou2024breaking}. Instead, SF110 has been extensively used to assess and improve \sbst{} tools like \evosuite{}, introducing a potential bias toward this latter category of approaches.
    \item Previous studies compare only \llmBased{} (mostly ChatGPT-3.5) and \sbst{} approaches.
\end{itemize}

\subsection{Overview}
Overall, we can conclude that a significant body of work is already on applying \llms{} in automatic test generation. Our work, however, aims to address several limitations and focus on aspects that are not covered by the existing body of work:
\begin{itemize}
    \item A study comparing the performance of different LLMs in the context of automatic test generation on a large-scale real-world benchmark with recent commits and code changes published after the release date of the \llms{} in our study.
    \item A comparison of the LLM-based test generation with search-based and
    symbolic execution-based methods.
    \item Analysis of strengths and weaknesses each test generation approaches regarding features of code under test.
\end{itemize}

\section{Test generation assessment pipeline}
\label{section:pipeline}

\begin{figure}
\centering
\includegraphics[width=\linewidth]{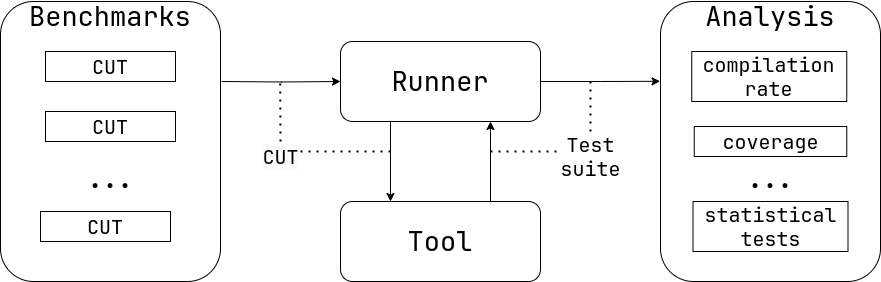}
\caption{Overview of the pipeline}
\label{fig:pipeline-overview}
\end{figure}

We have implemented a test generation pipeline to conduct an extensive study on the quality of automatic test generation for various methods. Figure~\ref{fig:pipeline-overview} shows the overview of the pipeline architecture.
The pipeline consists of four main parts: (1) benchmarks, (2) runner, (3) tools, and (4) analysis. The benchmark module provides a list of projects under test~(\projectUnderTest{}) in JSON format. Each \projectUnderTest{} description includes:
\begin{itemize}
    \item Name of the \projectUnderTest{};
    \item Version of \projectUnderTest{}~(commit hash);
    \item Path to the root of the project;
    \item Path to the sources directory;
    \item Path to the binaries directory;
    \item Full classpath;
    \item Name of the class under test~(\cut{}).
\end{itemize}

The runner handles interaction with the tools and saves the results of their work, which includes:
\begin{itemize}
    \item A path to the test sources directory;
    \item A list of fully qualified test names;
    \item A list of additional utility classes used by tests~(optional);
    \item A list of test dependencies~(e.g. JUnit, Mockito, etc.).
\end{itemize}

Each tool has a separate adapter interface that manages the interaction with the pipeline: process start, time limit, output handling, and result parsing. Currently, all the tools are executed as external CLI processes so that their execution doesn't affect the pipeline's core process. For extensibility, an adapter interface is the only requirement needed to add a new test case generation tool to our pipeline. The result of executing the runner is the test suite generated by the tool. The analysis module is executed as a post-processing. It extracts all the necessary metrics from the tools' outputs: compilation rate, coverage, mutation score, and test execution results. Then, it merges the tool's results with the pre-computed information about each benchmark~(will be covered more in Section~\ref{section:experimentalsetup}) and produces a CSV file with the final report. The pipeline is deployed as a swarm of Docker containers, cooperating via Docker Compose, making it easy to run in parallel.

\subsection{Dataset}
\gitbug{}~\cite{GitBugJava} is a reproducible benchmark of recent Java bugs published in 2024. It contains 199 bugs committed in 2023 extracted from 55 open-source repositories. This relatively new dataset has not been included in the training set for modern LLMs, which allows us to conduct more reliable experiments. \gitbug{} provides extensive information about each bug. In our study, we considered the following information: (1) project name, (2) repository URL, (3) commit hash of the buggy version, (4) commit hash of the patched version, and (5) bug patch.
%
Unfortunately, \gitbug{} does not provide explicit information about the buggy class that can be used as an automatic test generation target. Hence, for each bug, we extract information about the modified classes from the provided bug patch and manually analyze the source code to select one of them as a target~(in case of any conflicts).

All of the projects from \gitbug{} dataset were compiled using JDK 11, as it is the latest version of Java supported by \evosuite{}. However, some of the projects from the dataset require newer versions of Java. After excluding these cases, our final dataset contains 136 bugs from 24 open-source repositories. The dataset includes projects of various sizes ranging from 8,000 to 300,000 source lines of code~(\sloc{}) and \cuts{} of various complexity~(25--2,500 \sloc{}) with the average cyclomatic complexity score of $\approx 66$.

\subsection{Tools}

The core of our study is comparing test generation abilities of three main automatic test generation approaches: \sbst{}, \seBased{}, and \llmBased{}. Therefore, our pipeline integrates tools from each of these categories.
For SBST, we have selected \evosuite{}~\cite{evosuite} as it is the most well-known and proven automatic test generation tool for Java. It has shown the best results in recent editions of SBFT Java Tool Competition~\cite{SBFT23Competition, SBST22Competition, SBST2021}.
\kex{}~\cite{kex} is selected as a tool that implements \seBased{} approach to automatic test generation, as it is the only tool using this technique that participated in the latest SBFT Java Tool Competition~\cite{SBFT23Competition} and has shown good results in terms of coverage.
For the \llmBased{} test generation approach, we have selected \testspark{}~\cite{testSpark}. There are several reasons behind this choice:
\begin{itemize}
    \item Despite being an \intellij{} plugin, it can be easily executed in the command line by running the IDE in headless mode.
    \item \testspark{} supports multiple LLM models that can be easily interchanged~(fully integrated with three LLM platforms: OpenAI, HuggingFace, and JetBrains internal AI Assistant platform).
    \item \testspark{} uses the powerful code inspection of \intellij{} even in the headless mode, which brings flexibility to include different contexts in the prompt.
\end{itemize}

\subsection{Analysis}

\subsubsection{Execution-based metrics} These metrics are based on the compilation and execution of produced tests. 

\textbf{Compilation rate}. The compilation rate allows us to understand the tool's reliability for automatic test generation. This metric is especially important for \llmBased{} tools because they often struggle with producing correct code~\cite{LLMvsSBST}. However, most tools produce a test suite as a single source code file containing all the tests, and if there is even one syntax error, the whole test suite will not be correct. To handle these scenarios, we have extracted each test method into a separate source file. This allows us to record the compilation rate on a more granular level and also allows us to use correctly generated test methods for further analysis.

\textbf{Code coverage}. Coverage is one of the most used measurements of test suite quality~\cite{SBST2021}. We collected information about line, branch, and instruction coverage for each project using the JaCoCo~\cite{JaCoCo} library.

\textbf{Mutation score}. Mutation testing~\cite{MutationTesting} is a stronger test suite quality metric. Mutation score measures the ``strength'' of a test suite and characterizes its bug-discovering abilities. We use the PIT mutation testing system in the pipeline~\cite{Pit}.

\textbf{Failure reproduction}. As the \gitbug{} dataset contains a set of bugs, our pipeline evaluates the capabilities of these approaches in generating tests that can capture real-world bugs. For each entry in the dataset, we generate tests for the buggy version of the project. After, we execute tests on both buggy and patched versions of the PUT to record the differences. If there are any, it means that the tool can generate tests that are affected by the bug.

\subsubsection{Code feature metrics}
The quality of automatically generated tests is heavily dependent on the used approach and the code under test itself. Therefore, we perform static analysis on the \cuts{} in the benchmarks to extract information about distinct code features that may affect the quality of test generation. Based on that information, we can identify the strengths and weaknesses of our tools and approaches. For each \cut{}, we have collected the following information.

\textbf{Number of dependencies}. We define a dependency of \cut{} as an import used inside that \cut{}. Dependencies are categorized into internal, standard library, and external. Internal dependencies belong to the same project as \cut{}; standard libraries are built-in libraries in Java; external dependencies are third-party libraries not part of the previous two categories.

\textbf{Comments and Java docs}. Unlike the traditional automatic test generation approaches, \llmBased{} approaches use the source code of the \cut{} as the main input. Source code often includes additional contextual information in the form of comments and Java docs. These features can provide additional information about the program to \llm{} and thus affect the resulting tests. Additionally, we are interested not only in the presence of these features but also in the language that they are written in, as the initial instinct suggests that \llms{} should perform best in the English language.

\textbf{Condition types}. Branching points usually add a lot of complexity to the program. However, not every branching point is equal in its complexity. Therefore, we are interested to know what types of conditions have the most impact on the quality of generated tests. We have analyzed the source code of each \cut{} in the dataset and extracted the information about types and sources of variables in conditions.

\section{Experimental setup}
\label{section:experimentalsetup}

\subsection{Tool setup}

To address the nondeterministic nature of test generation algorithms used in this study, we repeated each execution 10 times as suggested by existing guidelines~\cite{GuideToStatisticalTests}. We also highlight that we applied the same methodology using multiple seeds and different sessions also for \llmBased{} approaches, addressing \textit{Limitation 3} discussed in Section~\ref{section:intro}.

We also need to define a time budget for our experiments as it can significantly influence the quality of generated tests~\cite{panichella2022test}. In our case, we set the time budget to 120 seconds, sufficient for the test generation tools to produce meaningful tests while remaining small enough to reflect practical, everyday-use scenarios (i.e., developers' need to receive quick feedback). Furthermore, this time limit aligns with the standards used in automatic test generation competitions like SBFT~\cite{SBFT23Competition, SBST22Competition}, where 120 seconds is one of the main categories.

However, most of the \llmBased{} approaches, including \testspark{}, are not designed to work within the time budget. \llmBased{} approaches cannot modify their test generation tactic regarding time limitations, as they depend on \llm{}'s performance: the model either successfully produces a compilable test or not. However, our preliminary experiments demonstrate that, on average, \testspark{} takes about 2-3 minutes on each \cut{}, which we consider comparable to \kex{}'s and \evosuite{}'s time budget.

Each tool we use in our experiments has many parameters that can affect its test-generation capabilities. Hence, we use each tool's default~(suggested) parameters as the default parameter values commonly used in the literature give reasonably
acceptable results~\cite{arcuri2013parameter} without incurring the additional computational cost required for parameter tuning.

We are using \kex{} version 0.0.8\footnote{\url{https://github.com/vorpal-research/kex/releases/tag/0.0.8}} in the concolic mode, and the only parameter we change is turning off built-in coverage computation. 
As for \evosuite{}, we are using version 1.2.0\footnote{\url{https://github.com/EvoSuite/evosuite/releases/tag/v1.2.0}}. Additionally, we are running \evosuite{} with DynaMOSA~\cite{DYNAMOSA} algorithm and with disabled runtime dependencies. We use DynaMOSA~\cite{DYNAMOSA} since it has been shown to outperform other evolutionary algorithms~\cite{campos2018empirical}, and was the default configuration in the SBFT competitions~\cite{SBST22Competition, SBFT23Competition}.

\llmBased{} test generation introduces two more major variables into the experiment setup: (1) model selection and (2) prompt engineering. These variables introduce the following questions into \llmBased{} test generation setup:
\begin{itemize}
    \item What model to choose?
    \item What information to include in the prompt?
    \item How to balance the prompt so it fits into the model's context?
\end{itemize}

As mentioned previously, \testspark{} allows seamless switching between different \llms{}. In this work, we selected the following models for the experiments: (1) \gpt{}, (2) \gpto{}, (3) \llama{}, and (4) \codellama{}.
This selection includes the most popular \llms{} today, and one code-specific model that will allow us to compare its performance to general-purpose ones.
%

Listing~\ref{lst:prompt} shows the default prompt used by \testspark{}. Initially, it includes the source code of the CUT~(\texttt{\$CODE}), signatures of the methods accessible during the test generation~(\texttt{\$METHODS}), and information about polymorphic relations~(\texttt{\$POLYMORPHISM}). However, if the prompt turns out too big for the used model, \testspark{} iteratively reduces its size by removing additional context information.

\begin{figure}
\begin{lstlisting}
Generate unit tests in Java for $NAME to achieve 100% line coverage for this class.
Dont use @Before and @After test methods.
Make tests as atomic as possible.
All tests should be for JUnit 4.
In case of mocking, use Mockito. But, do not use mocking for all tests.
Name all methods according to the template - [MethodUnderTest][Scenario]Test, and use only English letters.
The source code of class under test is as follows:
$CODE
$METHODS
$POLYMORPHISM
\end{lstlisting}
\caption{Default prompt used for \llmBased{} test generation}
\label{lst:prompt}
\end{figure}

We are using the latest version of \testspark{} that is available in its repository at the time of evaluation\footnote{\url{https://github.com/JetBrains-Research/TestSpark/tree/e6adea}}.

\subsection{Research questions}

The goal of our experiments is to answer the following research questions.
\begin{itemize}
    \item{\textbf{\rq{1}}}: \textit{What is the best \llm{} to use for automatic unit test generation?}
    \item{\textbf{\rq{2}}}: \textit{How do \llmBased{} automatic test generation approaches compare to traditional approaches on a large scale?}
    \item{\textbf{\rq{3}}}: \textit{What is the correlation between various qualities of code under test and the performance of different test generation techniques?}
\end{itemize}

To address \rq{1}, we focus on the execution-based metrics: compilation rate, line and branch coverage, and mutation score. \testspark{} is executed with each model under test~(\gpt{}, \gpto{}, \llama{}, \codellama{}) on the \gitbug{}. The performances are evaluated by analyzing the distributions of the results (metrics discussed in Section~\ref{section:evaluation}) over different runs using 
boxplots and descriptive statistics (median and mean). Additionally, we perform sound statistical analysis using the Mann-Whitney U test~\cite{MannWhitneyU} for the statistical significance, as it has already been used by prior work~\cite{SBFT23Competition}. We further complement our analysis with the Vargha Delaney $\hat{A}_{12}$ measure~\cite{VarghaDelaney} for effect size.
%
These statistical tests are computed pairwise for tools for each \cut{}. We use $p$-value threshold of $0.05$ to define significant differences and ``small'', ``medium'' and ``large'' magnitudes~\cite{VarghaDelaney} of the $\hat{A}_{12}$ statistics to determine the winner between the two tools in the comparison on each \cut{}. 

\rq{2} is addressed using all the same metrics mentioned for \rq{1}. This time, we are comparing the results of \kex{}, \evosuite{}, and the best performing \llm{} resulting from \rq{1} (which is \testsparkgpto{} as shown in Section~\ref{section:evaluation}) on the \gitbug{} dataset. We also used the same statistical test for this research question. In addition to the previously mentioned metrics, we also focus on the fault reproduction capabilities of the different approaches in this experiment. We execute the generated test suite on two versions of the project: the original buggy one and the patched one. Differences in the results between these two executions means that the tool was able to capture the fault in the buggy version of the project.

\rq{3} is addressed by performing correlation analysis~\cite{Correlation} between tool's performance and features of the \cut{}. \cut{} features are divided into two main categories:
\begin{itemize}
    \item Correlation between coverage metrics and \emph{code specific features} of \cut{}. \emph{Code specific features} include cyclomatic complexity, number of dependencies, presence of comments and Java Docs, and \sloc{}.
    \item Correlation between coverage metrics and \emph{branch condition types} in the CUT. \emph{Branch conditions} are divided into the following categories: primitives, \texttt{null} operations, \texttt{switch} conditions, type checks, static/global method/variable operations, string and regex operations, collections, and others.
\end{itemize}

Analyzing the correlation between listed features of \cut{} and the tool's performance on a large-scale benchmark of real-world projects allows us to better understand the strengths and weaknesses of the tools in different use cases. We use Spearman's rank correlation coefficient for correlation analysis~\cite{SpearmanCorrelation} as our data do not follow a normal distribution according to the Shapiro-Wilk test~\cite{shapiro1965analysis}.
\section{Evaluation and results}
\label{section:evaluation}

\subsection{\rq{1}: What is the best \llm{} to use for automatic unit test generation?}

\begin{figure}
\centering
\includegraphics[width=\linewidth]{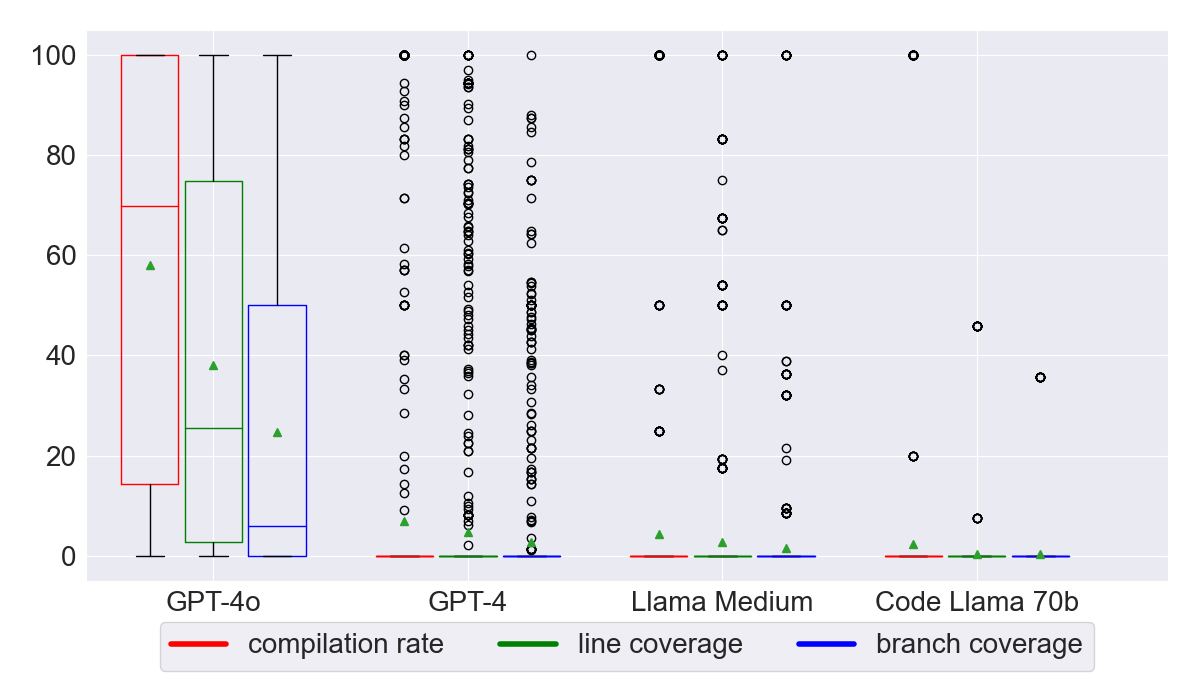}
\caption{Execution-metrics comparisons of the different \llms{} in \testspark{}}
\label{fig:llm-coverage}
\end{figure}

Figure~\ref{fig:llm-coverage} shows the average compilation rate, line, and branch coverage for four models: \gpto{}, \gpt{}, \llama{} and \codellama{}. According to these results, \gpto{} shows the best results on the benchmark, achieving an average $57.97\%$ compilation rate, $38.13\%$ line coverage, and $24.63\%$ branch coverage. Other models show worse results, achieving an average of less than $7\%$ compilation rate and $5\%$ line coverage.

\gpt{}, \llama{} and \codellama{} performed much worse than \gpto{}. This is due to the context limitations of the latter models: while \gpto{} has a 128k tokens context size, the others vary between 4-8k. Due to this, there are various instances where non-\gpto{} models failed even with the smallest possible prompt size~(i.e., only \cut{} source code). For example, out of 1360 total runs of \gpt{}, 1229 failed with the context size error.

In addition to the boxplots, pairwise comparison of the different \llms{} using statistical tests confirms the statistical superiority of \gpto{} on every measured metric. Out of $136 * 3 * 4 = 1632$ comparisons with other models, there are only two cases when \gpto{} achieves statistically lower performance metrics than another model.

Due to these limitations, we conclude that \gpto{} is the best viable option for automated test generation on real-world projects among the models investigated in this study. Therefore, we consider only \gpto{} in the remainder of our study and further experiments.

\subsection{\rq{2}: How do \llmBased{} automatic test generation approaches compare to traditional approaches on a large scale?}

\begin{figure}
\centering
\includegraphics[width=\linewidth]{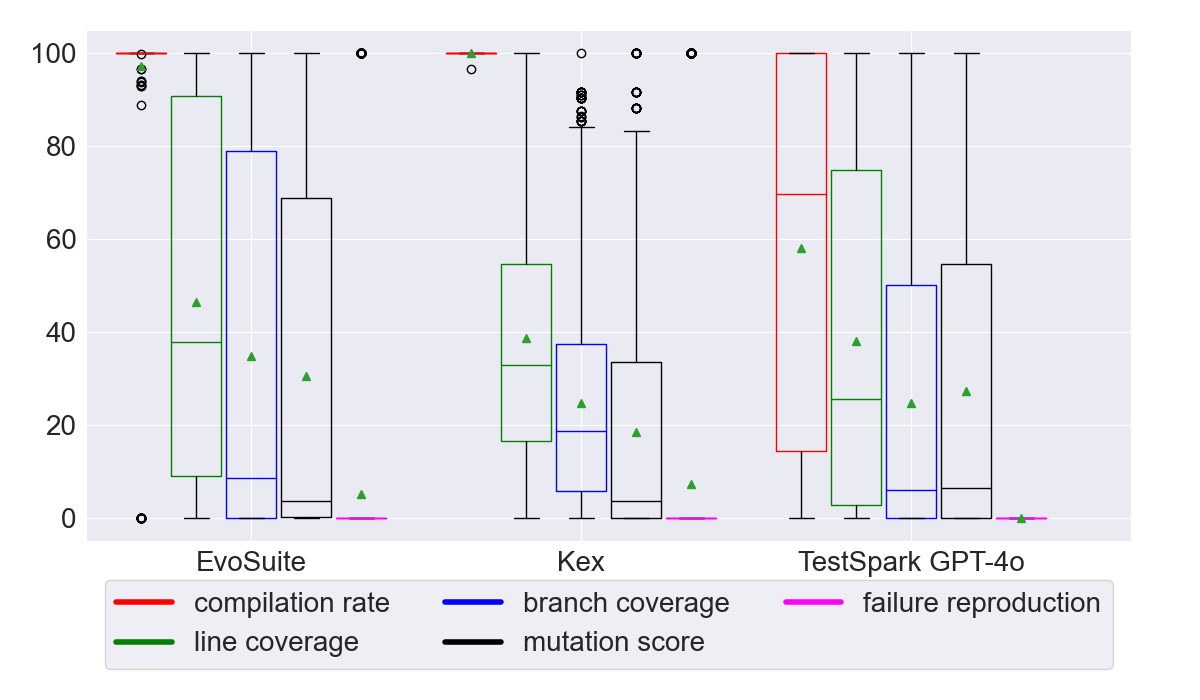}
\caption{Comparison of different automatic test generation tools}
\label{fig:tool-comparison}
\end{figure}

Figure~\ref{fig:tool-comparison} shows the comparison of three tools --- \evosuite{}, \kex{} and \testspark{} with \gpto{} model~(\testsparkgpto{} from now on) --- in terms of compilation rate, line and branch coverage, mutation score and failure reproduction rate. 

These graphs give us multiple insights into the performance of these tools. 
First, \kex{} has the best average compilation rate~(99.99\%) out of all three tools. Even though \seBased{} tools have full access to the classpath of the \cut{}, due to some implementation issues, \kex{} produced several uncompilable test cases. \evosuite{} is not far from \kex{} with a 97.30\% compilation rate, also caused by small implementation issues within this tool. \testsparkgpto{}, as mentioned previously, has a significantly lower compilation rate~(57.97\%) highlighting the unstable nature of \llmBased{} test generation.

In terms of coverage metrics, we can see that all three tools are close in terms of average line and branch coverage: 46.44\% and 34.91\% for \evosuite{}, 38.60\% and 24.72\% for \kex{}, 38.13\% and 24.63\% for \testsparkgpto{}. Although, \evosuite{} is slightly better, especially regarding average branch coverage. However, we can see that \kex{} is significantly better in terms of median branch coverage~(18.61\% versus ~4--6\% of \evosuite{} and \testsparkgpto{}). Additionally, we can see that \kex{} performs more consistently than the other two tools.

W.r.t. mutation score, \evosuite{} achieves the best average score~(30.56\%), while \testsparkgpto{} was the best in terms of median score~(6.32\%). While the mutation scores of \evosuite{} and \testsparkgpto{} are comparable to each other, \kex{}'s performance in this metric is noticeably worse as it sometimes generates wrong oracles.

All the tools performed poorly in terms of failure reproduction of original \gitbug{} bugs. \testsparkgpto{} did not reproduce any bug. \evosuite{} and \kex{} reproduced 5.88\% and 7.35\% of bugs, respectively. Statistical tests show that \kex{} performed better on 8 benchmarks, \evosuite{} performed better on 6, and both tools tied on two benchmarks.

\begin{figure*}[tbh]
    \centering
    \begin{subfigure}[b]{0.28\textwidth}
        \centering
        \includegraphics[width=\textwidth]{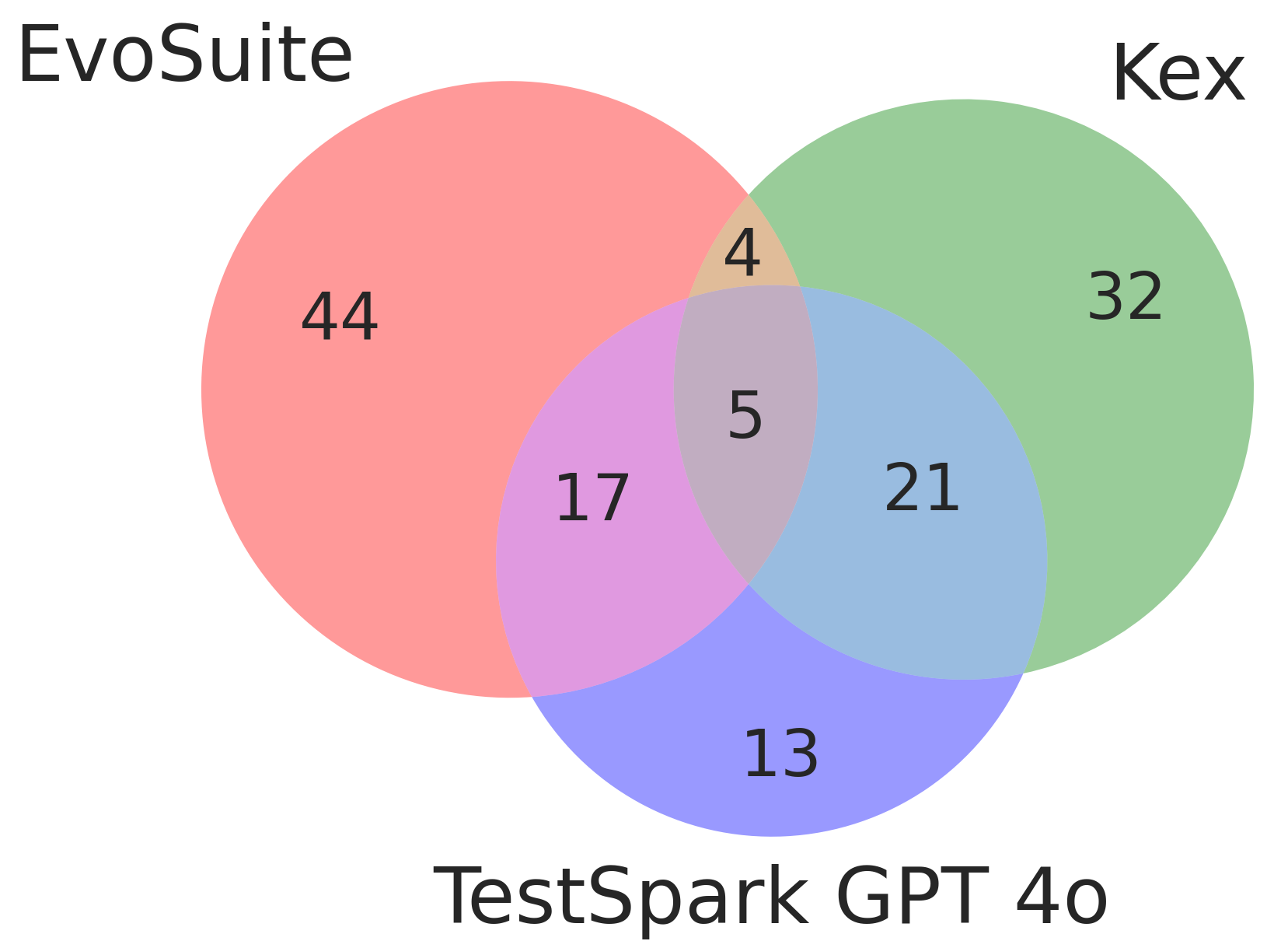}
        \caption{Line coverage}
        \label{fig:tool-venn-line}
    \end{subfigure}
    \hspace{10pt}
    \begin{subfigure}[b]{0.28\textwidth}
        \centering
        \includegraphics[width=\textwidth]{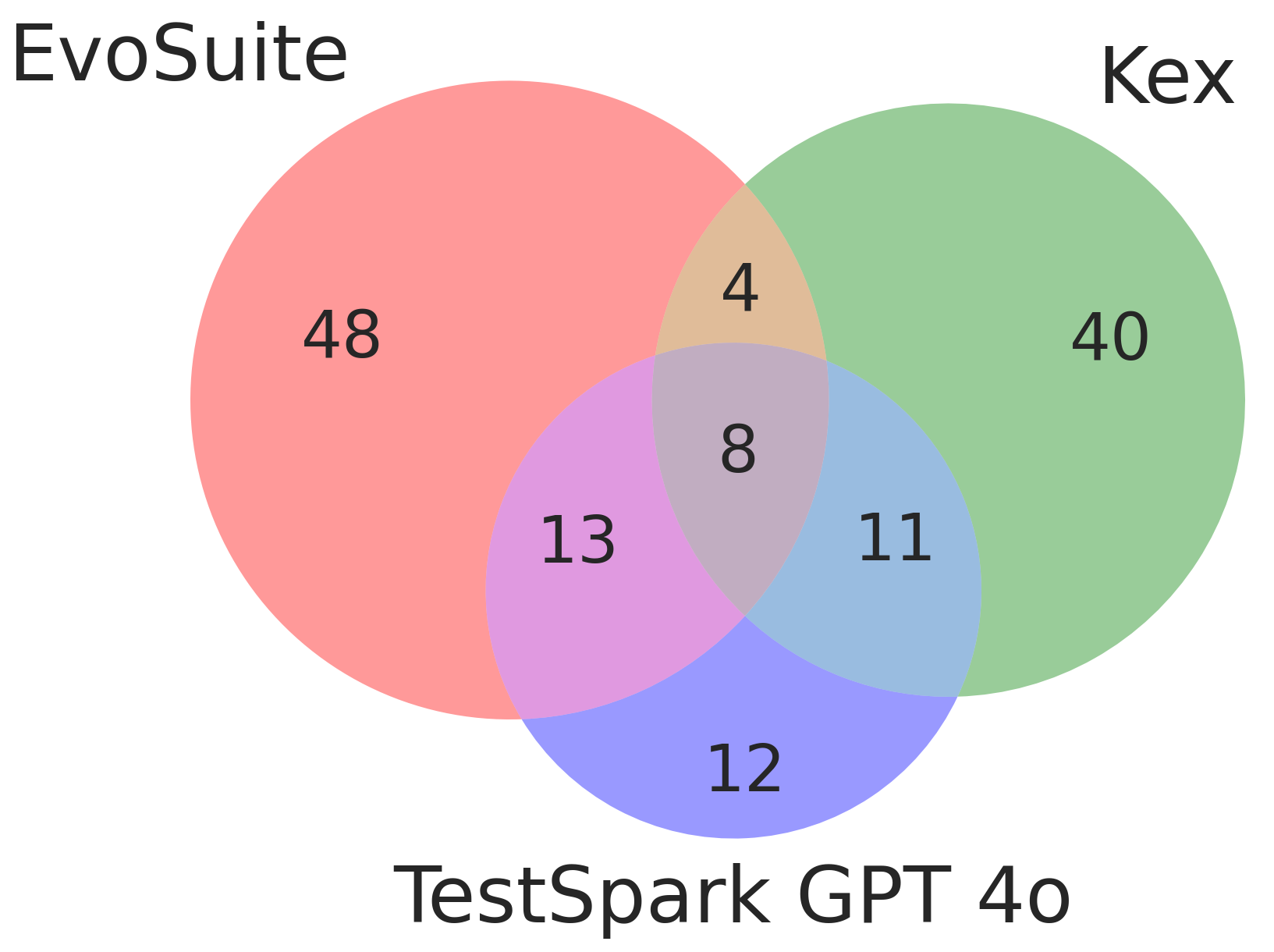}
        \caption{Branch coverage}
        \label{fig:tool-venn-branch}
    \end{subfigure}
    \hspace{10pt}
    \begin{subfigure}[b]{0.23\textwidth}
        \centering
        \includegraphics[width=\textwidth]{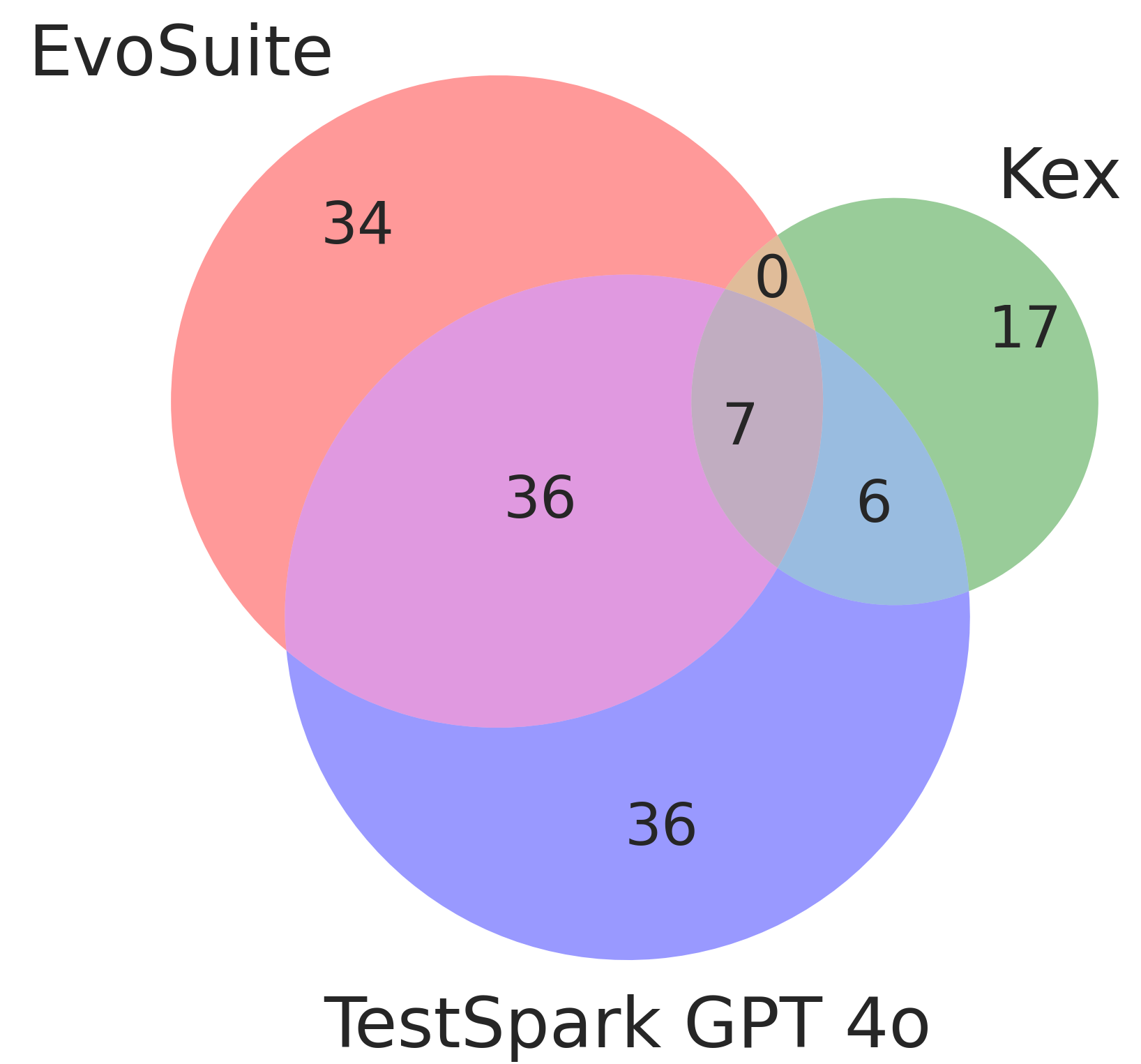}
        \caption{Mutation score}
        \label{fig:tool-venn-mutation}
    \end{subfigure}
    \caption{Venn diagrams of tools performances across \cut{}s, computed using pairwise comparisons of tools using $p$-value and effect size}
    \label{fig:tool-venn}
\end{figure*}

Additionally, for each \cut{}-metric pair, we conduct three statistical tests: \evosuite{} vs. \kex{}, \evosuite{} vs. \testsparkgpto{}, and \kex{} vs. \testsparkgpto{}. The winner between the two tools is determined based on $p$-value and effect size, with each tool earning 1 point for winning a statistical test. The tool(s) with the highest final scores for each \cut{}-metric pair are considered the winners. Figure~\ref{fig:tool-venn} visualizes these comparisons with Venn diagrams, categorizing each \cut{} based on the winning tool(s) in terms of line coverage~(\ref{fig:tool-venn-line}), branch coverage~(\ref{fig:tool-venn-branch}), and mutation score~(\ref{fig:tool-venn-mutation}). We can see that \evosuite{}{} performed best in terms of line and branch coverage, closely followed by \kex{}{}. \testsparkgpto{} falls behind in both metrics, but it performs noticeably better on the mutation score comparison, closely followed by \evosuite{}. \kex{} shows the worst performance on the mutation score.

In summary, we conclude that the tools are comparable regarding coverage-based metrics. \evosuite{} and \kex{} excel more at line and branch coverage, while \testsparkgpto{} shows better results in mutation score. W.r.t. to fault detection capability, the test generated by \evosuite{} and \kex{} outperformed \testsparkgpto{}, which failed to detect any fault.


\subsection{\rq{3}: What is the correlation between various qualities of code under test and the performance of different test generation techniques?}

\begin{figure*}[t]
\centering
    \begin{subfigure}{0.32\textwidth}
        \centering
        \includegraphics[width=\textwidth]{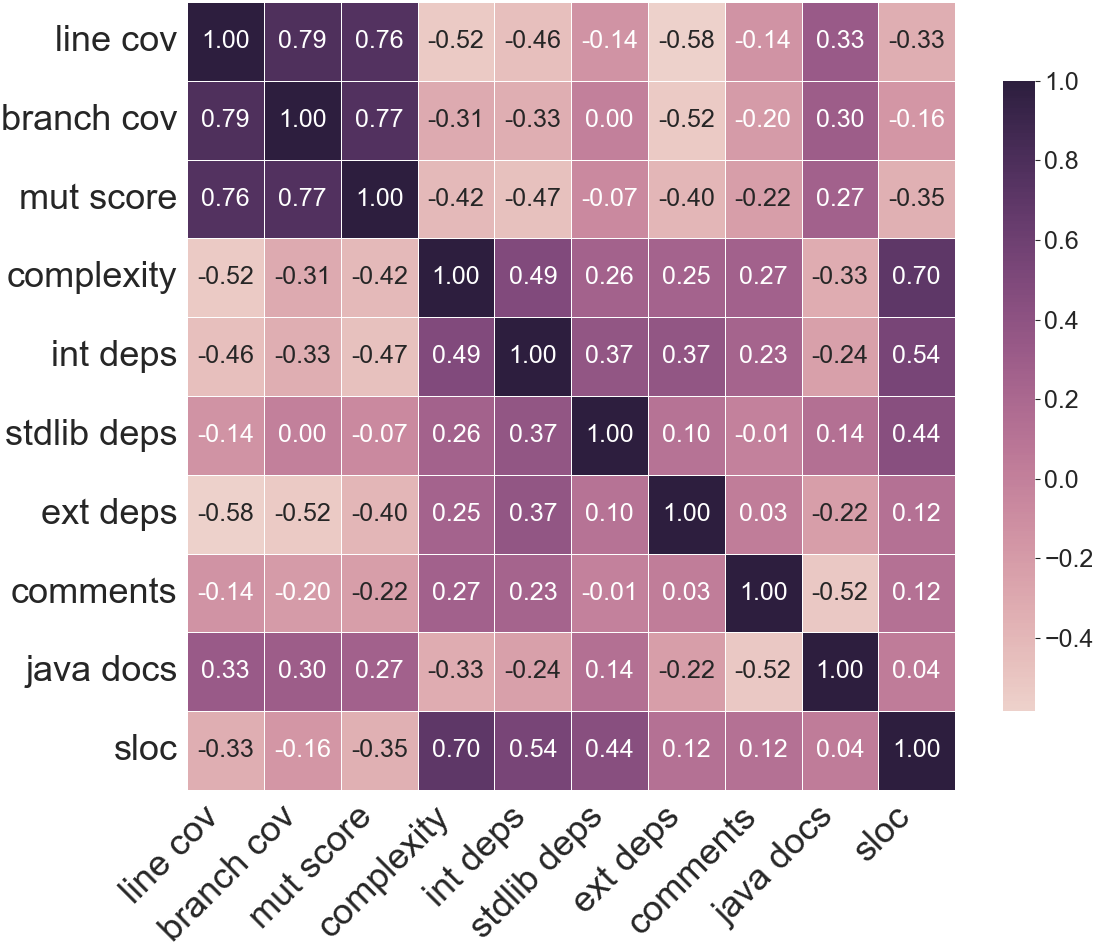}
        \caption{\evosuite{}}
        \label{fig:correlation-code-evosuite}
    \end{subfigure}
    \begin{subfigure}{0.32\textwidth}
        \centering
        \includegraphics[width=\textwidth]{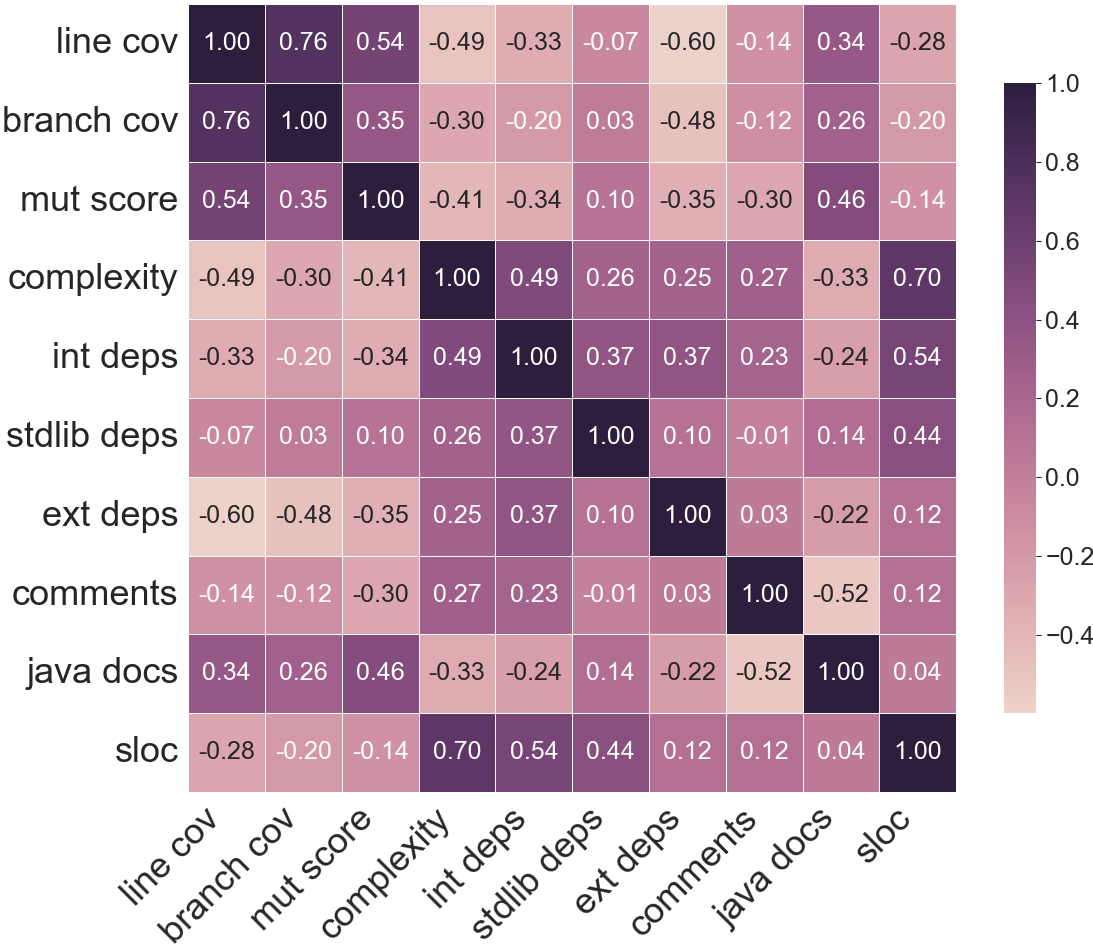}
        \caption{\kex{}}
        \label{fig:correlation-code-kex}
    \end{subfigure}
    \begin{subfigure}{0.32\textwidth}
        \centering
        \includegraphics[width=\textwidth]{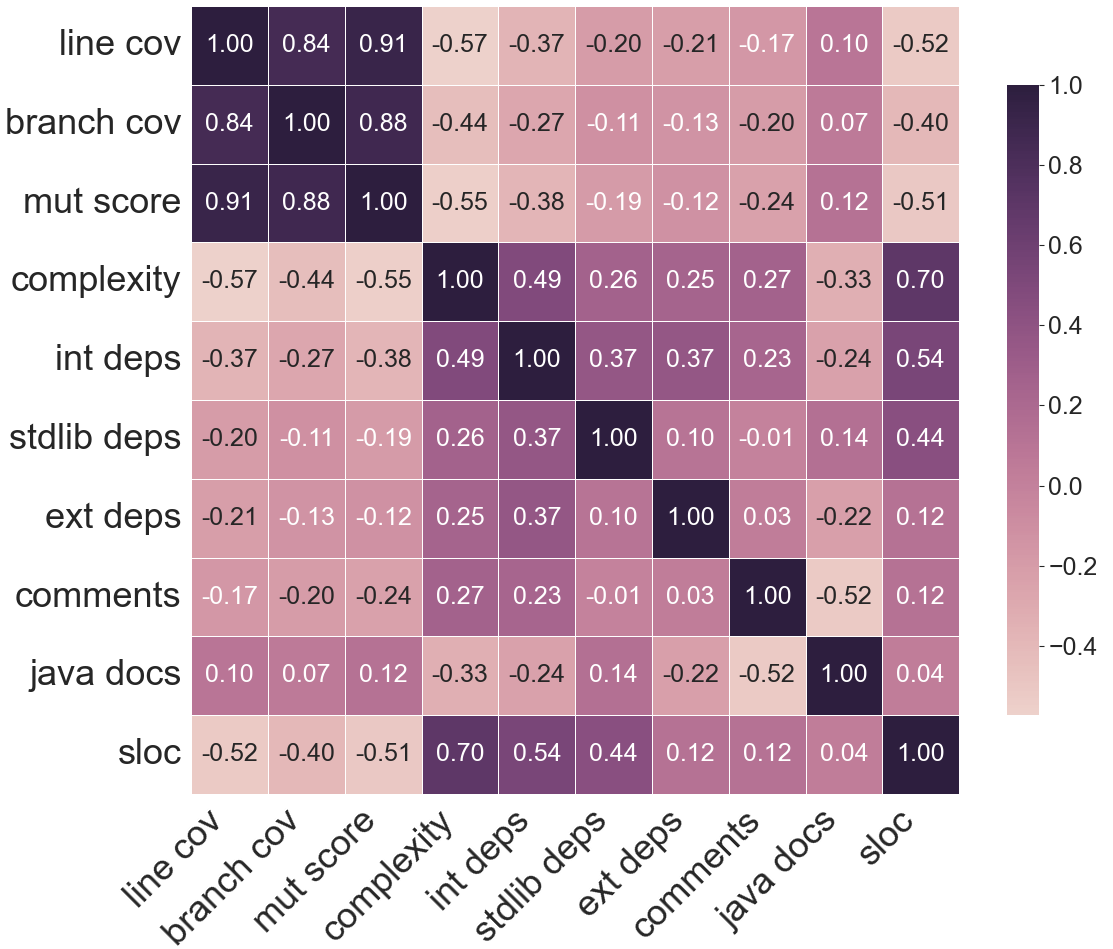}
        \caption{\testsparkgpto{}}
        \label{fig:correlation-code-gpt4o}
    \end{subfigure}
    \caption{Correlation analysis of tool performance and code-specific features}
    \label{fig:correlation-code}
\end{figure*}

Figure~\ref{fig:correlation-code} presents the results of the correlation analysis between the coverage-based metrics of the tools and the code-specific features of the \cut{}. We used Spearman's rank correlation coefficient to measure the strength and direction of these relationships. Correlations were classified as weak (0.00–0.30), moderate (0.31–0.60), or strong (0.61–1.00), with the sign indicating positive or negative correlations. These cutoff values follow the suggested classification guidelines~\cite{conover1999practical}.

We notice that all tools are significantly impacted by three features: the cyclomatic complexity of the \cut{}, the number of its internal dependencies, and \sloc{}. However, while the coverage of all the tools is affected approximately similarly by cyclomatic complexity, there is more variation in correlation with the number of dependencies and with \sloc{}. \kex{} shows the lowest correlation with these features, \evosuite{} shows a higher correlation with the number of internal dependencies, and \testsparkgpto{} shows a strong correlation with both the numbers of dependencies and \sloc{}. 

Surprisingly, all the tools show a weak correlation between their coverage performance and the number of Java standard library dependencies in the \cut{}, even though these are some of the most common dependencies in Java. Additionally, we can see that \evosuite{} and \kex{} have a strong negative correlation with the number of external dependencies of the project, while \testsparkgpto{} shows a weak correlation (i.e., lower than 30\%).

Additionally, \testsparkgpto{} shows a weak correlation with the presence of comments and Java Docs in the source code: comments seem to slightly worsen the test generation capabilities of \gpto{}~(-0.17), while Java Docs seems to have a weak positive correlation with the coverage-related metrics~(0.1). \kex{} and \evosuite{} do not correlate with these features. This result is expected since both tools work on the bytecode level (which does not include comments and Java Docs) and cannot access source code.

\begin{figure*}[t]
\centering
    \begin{subfigure}{0.32\textwidth}
        \centering
        \includegraphics[width=\textwidth]{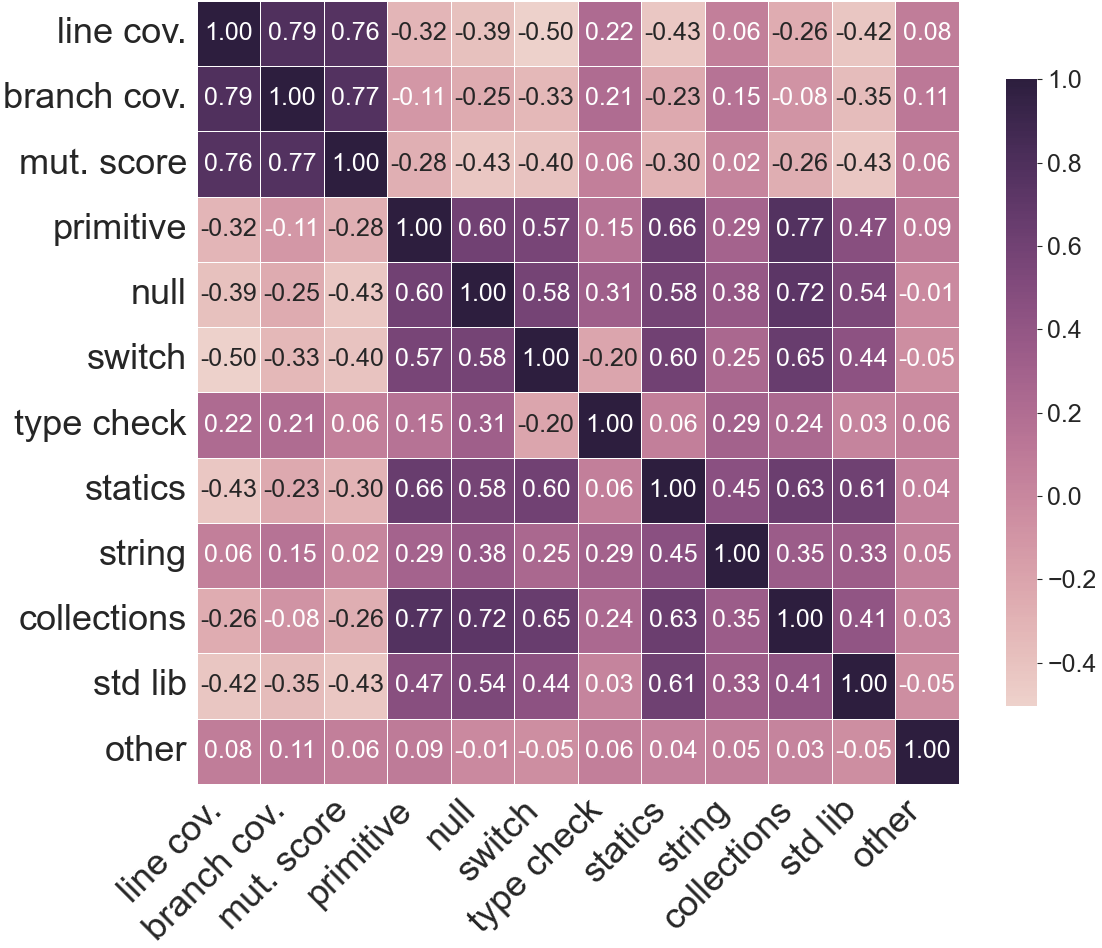}
        \caption{\evosuite{}}
        \label{fig:correlation-language-evosuite}
    \end{subfigure}
    \begin{subfigure}{0.32\textwidth}
        \centering
        \includegraphics[width=\textwidth]{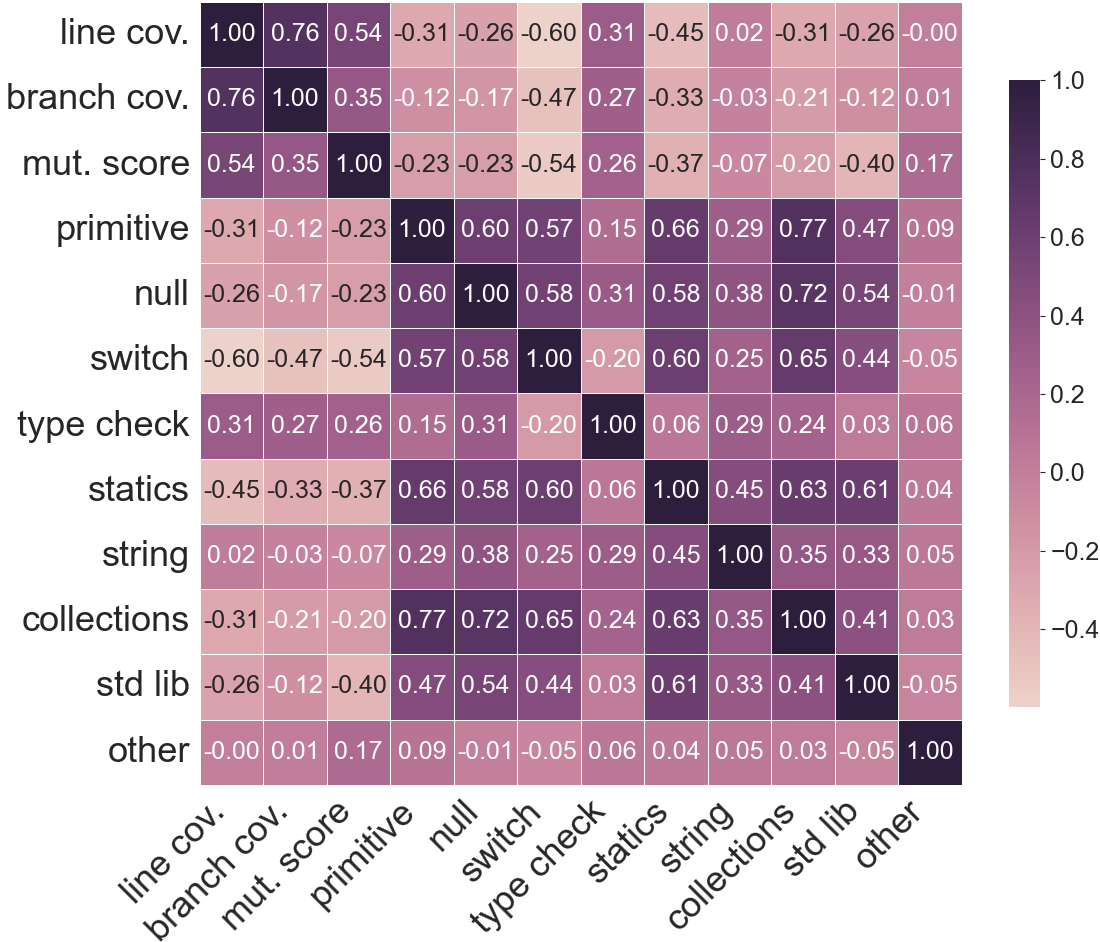}
        \caption{\kex{}}
        \label{fig:correlation-language-kex}
    \end{subfigure}
    \begin{subfigure}{0.32\textwidth}
        \centering
        \includegraphics[width=\textwidth]{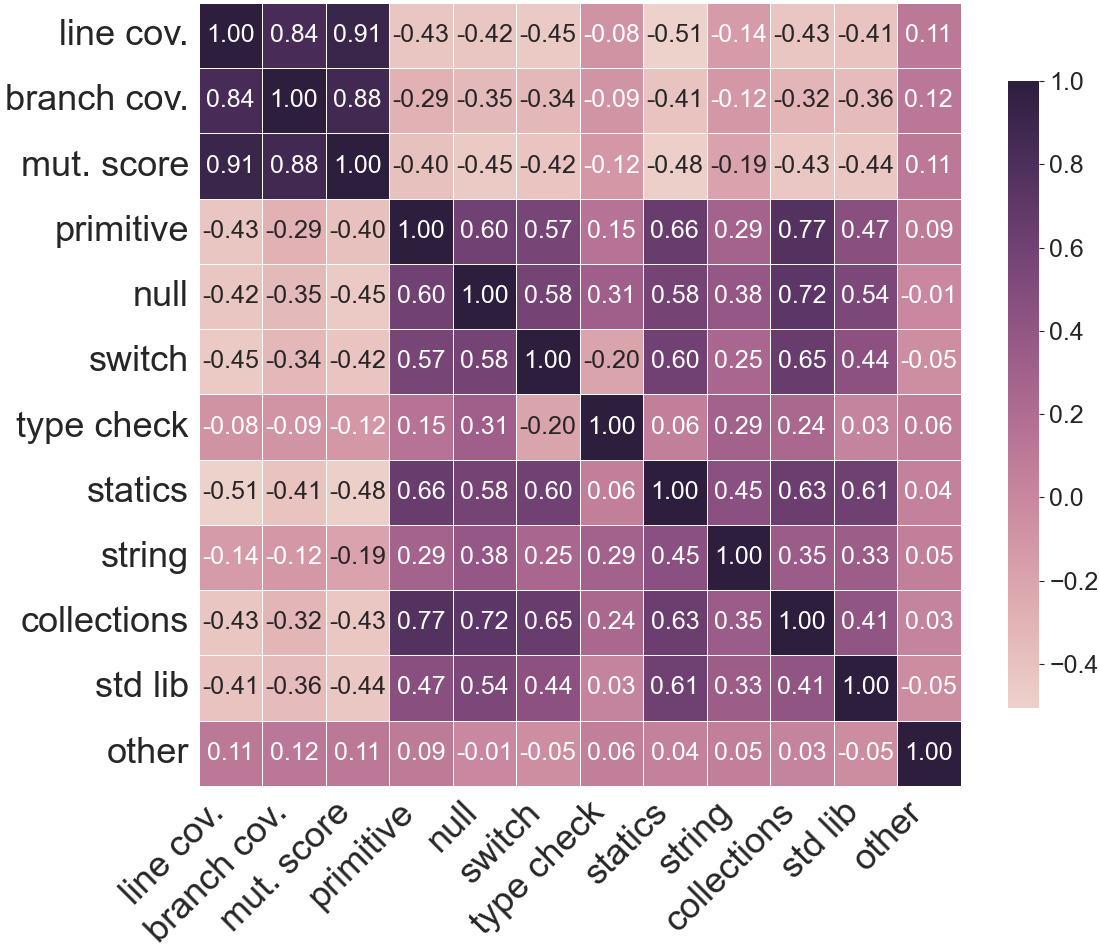}
        \caption{\testsparkgpto{}}
        \label{fig:correlation-language-gpt4o}
    \end{subfigure}
    \caption{Correlation analysis of tool performance and branch features}
    \label{fig:correlation-language}
\end{figure*}

Figure~\ref{fig:correlation-language} shows a correlation analysis between coverage-based metrics of the tools and branch types of the \cut{}. Overall, all the tools show almost no positive correlation with any of the branch types.
According to Figure~\ref{fig:correlation-language-evosuite}, \evosuite{} shows a negative correlation with several branch types. However, the switch conditions, static methods, and standard library calls negatively correlate to its performance. Additionally, we can see that coverage and mutation score achieved by \evosuite{} positively correlates with the number of type checks in the \cut{}. It is worth noticing that the presence of \texttt{null} checks has a moderate negative correlation with line coverage and mutation score. Null checks are a well-known issue in \sbst{} as they constitute the so-called flag problem~\cite{mcminn2004search, sbst}: an object is either null or not, and the existing heuristics (namely approach level~\cite{mcminn2004search} and branch distance~\cite{korel1990automated}) do not provide any guidance toward satisfying these checks.

Figure~\ref{fig:correlation-language-kex} shows \kex{}'s correlation analysis regarding coverage metrics. As \evosuite{}, \kex{}'s performance has a strong negative correlation with the number of switch statements and static calls in the \cut{}. Unlike \evosuite{}, \kex{} is much less affected by \texttt{null} checks, string operations, and standard library calls. Additionally, \kex{} shows a negative correlation with the number of primitive type conditions and collections in \cut{}. Moreover, we see that \kex{} has a strong positive correlation with the number of type conditions in the CUT.

Figure~\ref{fig:correlation-language-kex} presents the results of the correlation analysis of \testsparkgpto{}'s performance. Overall, \testsparkgpto{} shows a much higher negative correlation with most branch types encountered in the \cut{}.

\section{Discussion}
\label{section:discussion}


Regarding \rq{1}, the evaluation results give a definitive answer: \gpto{} is the best \llm{}~(among the ones we tested) for application in automatic test generation. As mentioned before, the main factor that played a role in this experiment is the context size of the model. Real-world programs are too big and too complex for the smaller \llms{}. Manual analysis of \cuts{} that were unaffected by context size limitations further shows that the smaller models perform slightly worse than \gpto{}. The three projects \texttt{traccar} (commit \texttt{074dc016d2}), \texttt{cbor-java} (commit \texttt{b010e8c62b}) and \texttt{semver4j} (commit \texttt{48ffbfd1f6}) are stand out as clear instances of the observed difference in performance. 

\rq{2} has no single definitive answer. \kex{} and \evosuite{} show very comparable performance on the line and branch coverage metrics, while \testsparkgpto{} falls behind. However, \testsparkgpto{} shows a better median (but not mean) mutation score than the other two tools. This seems to suggest \gpto{} might have a better understanding of the expected behavior of \cut{} and, therefore, potentially generate better oracles. However, this observation holds only for \gpto{} while the other \llms{} struggle to generate compilable test cases.

Finally, all tools struggle with fault detection capability. However, \testsparkgpto{} could not detect any fault in our experiments' runs. While the other two tools were able to seldom expose some faults in the benchmark.

Evaluation results demonstrate that each approach has a set of use cases where it outperforms others. Thus, the best possible automatic test generation tool should incorporate all these approaches to achieve better results. Our Venn diagrams highlight this complementarity.

As an example, the Venn diagrams~(figure~\ref{fig:tool-venn}) demonstrate that \evosuite{} and \kex{} have a very small intersection across all three metrics, meaning that these two tools~(and, therefore, approaches) excel at different \cuts{}. Moreover, \kex{} has an even smaller intersection with both tools in the mutation scores, meaning that, even though it generates worse oracles in general, in some cases, its oracles are unique in comparison with \evosuite{} and \testsparkgpto{}.

\textit{Manual analysis}. 
We manually analyzed the subset of 32 benchmarks that demonstrated significant differences in tool performances. Although it is hard to generalize beyond our benchmark and dataset, we can provide valuable insights:
\begin{itemize}
    \item \evosuite{} and \testsparkgpto{} are very effective at generating tests for string-processing \cuts{}. For example, benchmark \texttt{jsoup} (commit \texttt{e1880ad73e}) is a URL builder class; \evosuite{} and \testsparkgpto{} are able to generate valid URL strings in their tests. However, when it comes to more complex and non-standard formats, \testsparkgpto{} can outperform \evosuite{}~(e.g., the \texttt{semver4j} commit \texttt{48ffbfd1f6}).
    \item \testsparkgpto{} is often let down by the compilation errors of the generated tests. Benchmarks like \texttt{Simple-DSL} (commit \texttt{81182e58bd}) and \texttt{traccar} (commit \texttt{28440b7726}) show that when all the tests compile, \testsparkgpto{} can achieve on par or even better coverage than \kex{} and \evosuite{}.
    \item \evosuite{} and \testsparkgpto{} use mocking when working with interfaces and standard library classes. \kex{} only uses mocks when working with abstract classes. Benchmarks like \texttt{dataframe-ec} (commit \texttt{1109752c4e}) and \texttt{traccar} (commit \texttt{596036dc33}) show how mocking allows \evosuite{} and \testsparkgpto{} outperform \kex{}.
    \item Our manual analysis confirmed that \testspark{} with \gpto{} generates more and better oracles in many cases. For example, on \texttt{traccar} (commit \texttt{596036dc33}), both \evosuite{} and \kex{} failed to generate oracles for some of their tests.
    \item \kex{} often outperforms other tools on a very complex \cuts{} with the complex arguments. Benchmarks like \texttt{graphql-java-annotations} (commit \texttt{6d9d7a79de}), \texttt{jsoup} (\texttt{b6f652cef6}), and \texttt{frigga} (commit \texttt{6b520bbb2e}) highlight that. Additionally, \kex{} outperforms other tools on benchmarks that contain some bit-level operations and conditions~(e.g. \texttt{traccar-2be2a4558a}).
    \item Our results suggest that the 120-second time limit is not always enough for the intensive approach of \kex{}. Benchmarks \texttt{java-solutions-7a73ea56d0} and \texttt{cbor-java-b010e8c62b} are instances that led us to this conclusion.
\end{itemize}

\textit{Correlation analysis}. 
Feature-based correlation analysis also highlights several insights into automatic test generation tools' performance. First, each tool demonstrates moderate dependency on the \cuts{} \sloc{} and computational complexity. Interestingly, \testsparkgpto{} is more dependent on the \sloc{} than \kex{} and \evosuite{}, meaning that \llmBased{} models are better for smaller classes not only because of the context limitations. Secondly, the correlation analysis demonstrates that all the tools have a moderate correlation with the number of internal and external dependencies of \cut{}. \testsparkgpto{} is more affected by internal dependencies, meaning that it is hard for \llmBased{} methods to work with closely interconnected projects. \evosuite{} and \kex{} are expectedly more affected by the number of external dependencies, as their number heavily affects the complexity of the analysis.

We also focus on evaluating how the additional information within the code~(like comments and Java Docs) can affect the \llms{} performance. We do not see any strong correlation between these features and the tool's performance, but still, we can see that comments in code have a weak negative correlation~(-0.17), and Java Docs demonstrate a weak positive correlation~(0.1). Additionally, it is worth noting that 135 of 136 projects used the English language in the code for variable and class names, comments, docs, etc. In additional experiments, where we translated the \cuts{} to Spanish using \gpto{}, no significant performance variations were observed across different languages. Due to space constraints, further details of this analysis are not included in the paper.

Branch type-based correlation analysis of the tool's results also demonstrates some interesting results. Firstly, all tools demonstrate a negative correlation with almost all branching types, meaning that, expectedly, any branch complicates things for the test generation. Secondly, all tools demonstrate a moderate negative correlation with the number of primitive and switch conditions in the \cut{}. Although it is unexpected at first, manual analysis of the results shows that it is mainly because of the following:
\begin{itemize}
    \item Sheer number of such conditions; especially switch cases usually contain many branching points;
    \item The source of the variables in these conditions: in many cases, these variables are products of some complicated operations~(e.g., file reads, collection interactions, etc.).
\end{itemize}

Third, we can see that \testsparkgpto{} shows a moderate negative correlation with most branch types except strings and type checks. Manual analysis results confirm that trend: \llmBased{} test generation shows promising results with string processing \cuts{} and struggles more with the others.

Finally, only \kex{} and \evosuite{} show a weak positive correlation with the number of type checks in the code. It is the only condition type that positively correlates with any tool's performance. While not obvious, this fact can be easily explained. Type checks are usually relatively easy for these tools~(especially for \kex{}), as they are easily satisfiable and do not have a lot of options. At the same time, type checks provide the tool with additional contextual information and simplify the further generation.
\section{Threats to validity}
\label{section:threats}

In this section, we acknowledge the threats that may affect the validity of our experimental results. We can identify three main groups of threats:
\begin{itemize}
    \item \textit{internal} --- threats that are rooted in our implementation;
    \item \textit{external} --- threats related to the resources not directly connected to our implementation;
    \item \textit{conclusion} --- threats related to the reliability of conclusions drawn from our experimental results.
\end{itemize}

\paragraph*{Internal threats}

We ensured the correctness of every implementation step and manually analyzed the portions of the result to ensure their correctness. However, it is still possible that our implementation contains some bugs and issues.

Additionally, the difference in the time budget handling for \llmBased{} and traditional test generation approaches can introduce threats to the validity of our results. It is hard to compare these approaches fairly because of their unique features. Our experiment results suggest that the time budget handling did not substantially affect the quality of our results.

\paragraph*{External threats}

\llms{} introduces many data-related threats to the computer science research. Data leakage is one of the most important ones. Because of the nature of \llm{} training, modern \llms{} have been exposed to many open-source software projects during the training process. Thus, they may demonstrate significant differences in test generation quality on the projects they have seen during (pre)training and those they have not seen. Even though \gitbug{} dataset was published later than the alleged training time of all the \llms{} used in our experiments, the projects used in \gitbug{} were still available for \llms{} for (pre)training.

Another potential threat is the dataset itself. It contains a lot of different projects and bugs; however, it may not be representative enough. Moreover, the \gitbug{} dataset has a small imbalance in the project distribution. Out of the 136 bugs used in our evaluation, 29 were related to \texttt{jsoup} project, and 70 were related to \texttt{traccar} project.

\paragraph*{Conclusion threats}
A potential threat to validity is related to the non-determinism of \llms{}, \evosuite{}, and \kex{}. To address this potential threat (and the limitation of existing comparative studies involving \llms{}), we run each tool ten times of each \cut{} in our benchmark. Besides, for \llms{}, we use different/separate sessions for promoting/queries to ensure that the model did not learn from past interactions and prompts. For our analysis, we compare the results based on the median and mean results achieved w.r.t. well-established test quality metrics (i.e., coverage, mutation score, and fault detection capability) and relied on sound statistical analysis, following existing guideliens~\cite{GuideToStatisticalTests, sallou2024breaking}. More specifically, we have used non-parametric tests that do not make any assumption
on the data distributions being compared, namely the Mann-Whitney U Tests (or the Wilcoxon rank sum test), the Vargha-Delaney $\hat{A}_{12}$ statistics, and the Spearman's rank correlation coefficient.

\section{Conclusion}
\label{section:conclusion}

In this paper, we present the results of our extensive evaluation of three automatic test generation approaches: \sbst{}, \se{}, and \llmBased{} test generation. We evaluated three tools that implement the aforementioned approaches --- \evosuite{}, \kex, and \testspark{} respectively ---  on the bugs from \gitbug{} dataset with the main focus of highlighting the strengths and weaknesses of each approach.

Our evaluation results demonstrate that \llmBased{} test generation approaches are very heavily reliant on the context size of the \llm{}: 4-8k context size is not big enough for real-world applications. The best performing \llm{} out of the four tested is \gpto{}.

\evosuite{} and \kex{} perform very close in terms of coverage-based metrics like line and branch coverage, while \testsparkgpto{} falls slightly behind. However, the \llm-based {} approach shows a noticeable improvement in the mutation score, suggesting that \llms{} are more capable of deeper code understanding. Our evaluation also revealed that \llmBased{} performed worse than traditional approaches in terms of fault detection capabilities and failed to reproduce anything. Manual analysis results additionally highlighted some of the differences in tool performances.

We plan to create a new multi-approach test generation tool in future work. Our evaluation results suggest that we can find the best approach for each \cut{} and combine the strengths of different approaches.

\section{Data availability}

The reproduction package with the dataset collection scripts and the results presented in the evaluation is available at \url{https://doi.org/10.5281/zenodo.13862019}. Reproduction package does not include the source code of the pipeline due to anonymity issues. However, the link to the pipeline repository will be published after the double-blind review process.

\section{Acknowledgements}

This work was conducted as part of the AI for Software Engineering~(AI4SE) collaboration between JetBrains and Delft University of Technology. The authors gratefully acknowledge the financial support provided by JetBrains, which made this research possible.

\bibliographystyle{IEEEtrans}
\bibliography{main}

\end{document}